# On the role of moisture in triggering out-of-plane displacement in paper: from the network level to the macroscopic scale.

E. Bosco[a,*], R.H.J.Peerlings[b], B.A.G. Lomans[b], C.G. van der Sman[b], M.G.D. Geers[b]

[a] *Department of the Built Environment, Eindhoven University of Technology, P.O. Box 513, 5600 MB Eindhoven, The Netherlands*

[b] *Department of Mechanical Engineering, Eindhoven University of Technology, P.O. Box 513, 5600 MB Eindhoven, The Netherlands*

## Abstract

The response of paper to humidity variations is a complex, inherently multi-scale problem. The hygroscopic swelling of individual fibres and their interactions within the fibrous network govern the macroscopic, sheet-level response. At this scale, moisture induced instabilities and out-of-plane deformations may occur, which are critical for a number of industrial applications. This work specifically focuses on several aspects of this important issue. A macroscopic phenomenological hygro-mechanical model is first proposed, which aims at predicting moisture induced out-of-plane deformations in paper sheets. The constitutive model is based on the relation between these deformations and typical irreversible phenomena associated to the history of paper manufacturing, i.e. the release of dried-in strains. The model is used to describe bending induced by moisture gradients through the thickness of the sheet as well as buckling due to moisture variation in the presence of mechanical constraints. The results of the model show that the anisotropic sheet-level hygro-expansion has a strong influence on the instability phenomena. Moreover, a comparison with experiments provides adequate semi-quantitative estimates. An additional step is made towards the multi-scale understanding of paper hygro-mechanics. The fundamental physical mechanisms governing the macroscopic moisture induced response are investigated on the basis of the underlying fibrous network. To this aim, a meso-structural model is developed which consists of a network of fibres randomly positioned in a planar region according to an orientation probability density function. A series of network simulations reveals that upon moisture content variations the expansion of the inter-fibre bonding regions essentially drives the overall deformation. Particularly in the case of anisotropic fibre orientation, this explains the origin of the macro-scale anisotropic hygro-expansion, which is essential for the observed sheet-level instability phenomena.

## 1. Introduction

Significant dimensional variations occur in paper sheets when they are subjected to changes in moisture content. An inhomogeneous moisture distribution both in the plane of the sheet and through the thickness may result in instabilities and out-of plane displacements [1, 2, 3], which are possibly irreversible due to the release of the dried-in internal strains induced during the manufacturing process. The moisture related dimensional stability of paper is critical in several industrial applications, for instance in the field of digital ink-jet printing [4]. A simple illustrative example, which idealizes the ink-jet or off-set printing process, is proposed to represent the phenomena of interest. Figure 1(a) shows two paper strips, whose length direction is perpendicular to the machine direction (MD). In this direction, i.e. the cross direction (CD), the sheet is dried freely and there are no initially dried-in internal strains. The samples are subjected to a moisture cycle by wetting their top surfaces. During the experiment, one strip of the paper is placed on a solid surface, which prevents bending; the other strip is free to bend out-of-plane. In the wetting stage, the moisture content variation in the top regions induces expansion, whereas the bottom regions remain dry. This results in bending of the free paper sample, and in the "drying-in" of internal strain (and stress) in the constrained sample, see Figure 1(b). After drying, the free standing paper strip returns approximatively to its original configuration. The constraint is removed for the second sample. The built-in internal stress relaxes, leading to a significant out-of-plane deformation, as illustrated in Figure 1(c).

*Corresponding author. Tel.: +31-40-247-3257
*Email address:* e.bosco@tue.nl (E. Bosco)

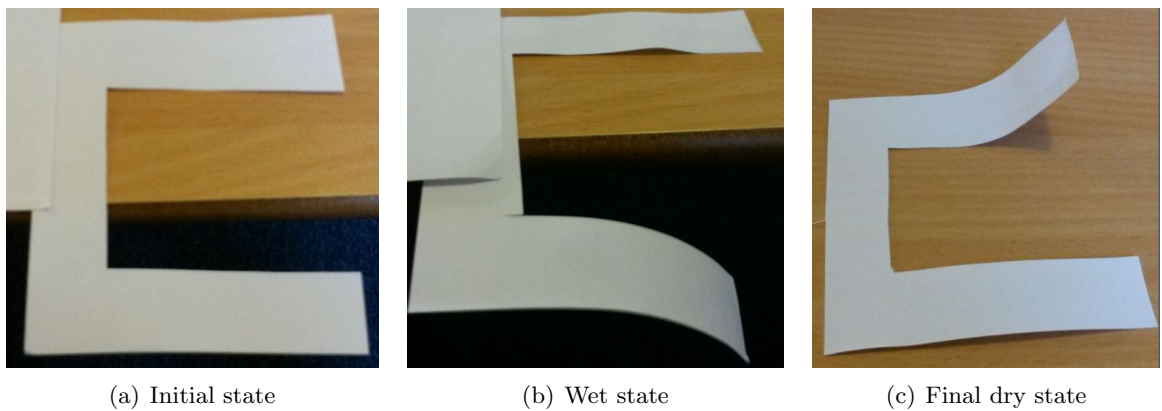

(a) Initial state  (b) Wet state  (c) Final dry state

Figure 1: Experiment showing out-of-plane displacement of paper subjected to moisture gradients through the thickness.

The sheet level moisture induced deformations are ultimately governed by the hygro-expansive response of individual fibres, which is transferred through inter-fibre bonds within the fibrous network [5, 6]. The swelling of a single fibre is highly anisotropic and, as a consequence, complex interactions between mechanical and hygro-expansive properties take place in the bonding areas, affecting the overall material response. Additional complexity is given by the fact that the fibre orientation distribution is also anisotropic [7]. The relation between phenomena at the fibre- and network- scale and the effective behaviour of paper is therefore crucial.

In this complex scenario, the present work aims on the one hand to develop an effective macro-scale model that is able to predict the out-of-plane moisture induced displacement in paper sheets, for different moisture paths and boundary conditions. On the other hand, it focuses on identifying the relevant mechanisms at the level of the fibrous network that govern the resulting macroscopic behaviour.

In the first part of the paper, a phenomenological model for the hygro-mechanical behaviour of paper is presented. Several contributions in the literature describe the non-linear response of paper under mechanical loading, either with elasto-plastic [8, 9, 10, 11, 12], visco-elastic [13, 14], or visco-elasto-plastic models [15, 16]. Moisture related effects have been included in [17, 18, 19].

The key feature of the present model, departing from [20], is the fact that it focuses on the description of moisture induced out-of-plane deformations and particularly on their relation with the release of dried-in strains. The constitutive model is based on kinematic hardening plasticity, with moisture dependent mechanical properties. These model ingredients allow to represent the presence of dried-in strains and their release during a moisture cycle. The developed model is formulated on a discrete basis, i.e. paper sheets are represented through a planar network of beams, whose properties are a function of the beam orientation, either in machine direction or in cross direction. This naturally provides an anisotropic response. Note that moisture induced instabilities are related to the relaxation of stresses that arise in the plane of the sheet; this justifies the assumed two dimensional description. This aspect is illustrated through a numerical/experimental example in which a buckling type of deformation occurs due to a uniform moisture variation in the presence of an in-plane mechanical constraint. An inhomogeneous moisture variation through the thickness can be another cause of out-of-plane deformation. This is described in the proposed approach by modelling the thickness of the sheet through a vertical stack of similar elasto-plastic uni-axial elements, with constant material properties but a different moisture content in the thickness direction. Two numerical simulations explore how moisture gradients through the thickness trigger a bending type of deformation, including the analysis of additional, non trivial effects depending on the kinematical expansion conditions (free or constrained).

The second part of this work investigates the fundamental meso-structural mechanisms induced in the fibrous network by hygro-expansion and their influence on the macro-scale response. This is done by developing a two dimensional representative volume element (RVE) of the fibrous network, in which straight fibres are randomly positioned in the plane following a certain orientation distribution, similarly to e.g. [21, 22, 23, 24, 25]. Additional focus is put here on hygro-expansive effects by modelling the fibres as two dimensional domains. Both longitudinal and transverse hygro-mechanical properties are included. Their interplay in the bonding areas largely governs the effective response, particularly influencing the hygro-expansive anisotropic behaviour, which is essential for the observed/computed macro-scale response. A series of network simulations is performed, which illustrate the occurrence of this phenomenon. The problem is restricted to hygro-elastic behaviour; elasto-plastic effects can be included following the approach presented in [26]. Note finally that, whereas the main focus here is on the local network response, structure-property relations may be also formulated, which allow to predict the sheet-level hygro-expansivity as a function of meso-structural parameters such as the single fibre properties and fibre orientation distribution. Advances in this direction have been made in [27, 28].

The paper is organized as follows. In Section 2, the macro-scale phenomenological model is proposed. Numerical examples exploring buckling and bending type of out-of-plane deformations are examined in Section 3. The network model and the corresponding meso-structural observations are discussed in Section 4. Conclusions are finally given in Section 5.

## 2. Macro-scale phenomenological model

*2.1. Typical free expansion response*

The macro-scale behaviour of paper during variations in moisture content is strongly influenced by the drying conditions in the manufacturing process. The hygro-expansive response due to a repeated moisture cycle in a machine produced sheet in the machine direction and in the cross direction is illustrated in Figure 2. In MD, the paper is subjected to web tension during production; in the drying phase of this stage internal stresses and strains are dried-in. These stresses and strains can be released when paper is next moistened for the first time. The release of dried-in strains induces an irreversible shrinkage. After the first moisture cycle, a linear reversible response is generally retrieved. This is illustrated in Figure 2(a). A different behaviour occurs in the cross direction, along which the degree of constraint during production is generally less. Hence, during (re)wetting, the irreversible strain release is substantially lower, see e.g. [29]. For this reason, the hypothesis is made here that in the cross direction the material behaves as if it has been dried freely. This implies that no irreversible strain release takes place and a linear relation between the hygro-expansive strain and moisture content can be assumed, as shown in Figure 2(b). Moreover, the hygro-expansive coefficient, denoting the proportionality ratio between the reversible strain variation and the moisture content variation, is smaller in the machine direction than in the cross direction, i.e. $\overline{\beta}_{MD} < \overline{\beta}_{CD}$ [6, 30].

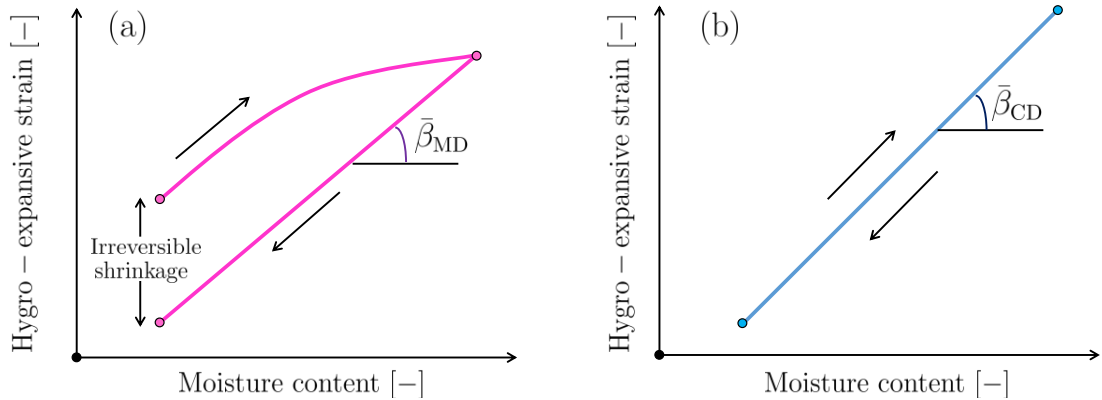

Figure 2: Sketch of the assumed hygro-expansive strain as a function of moisture content during a wetting-drying cycle for a machine made paper sheet in (a) machine direction and (b) cross direction, with $\overline{\beta}_{MD} < \overline{\beta}_{CD}$.

2.2. Geometry

Although paper at the macro-scale is considered as a continuous material, it is here idealized as a discrete system. A two dimensional network of beams is used to model the in-plane structure of paper, as illustrated in Figure 3(b). The constitutive description is therefore formulated on the basis of a uni-axial element, with different properties in the two in-plane principal directions. A two dimensional description is used, as instabilities are generally triggered by stresses in the plane. The same constitutive model can be used to model bending phenomena due to moisture gradients through the thickness. In this case, the uni-axial element is repeated vertically, see Figure 3(c), assuming constant properties in the thickness direction (ZD). The adopted hypotheses allow to formulate a discrete structural model that can be implemented in a simpler way than a continuum formulation, using standard commercial software resources.

2.3. Constitutive model

The constitutive model proposed in this work is based on the characteristic hygro-expansive response of paper shown in Figure 2. Despite the fact that paper based materials are generally characterized by a visco-elastic behaviour [13, 16, 17, 18], temporal effects are here neglected. This is justified by the fact that the time scale involved in the moisture cycle considered is sufficiently large to relax all the time dependent deformations. A small-strain elasto-plastic constitutive model with linear kinematic hardening is therefore considered, in which the elastic moduli, hardening moduli and yield stress are functions of the moisture content. This allows to model the dried-in strain that is initially stored in the paper in machine direction due to the production process. Moreover, a moisture dependent yield stress enables to describe the release of this dried-in strain upon wetting by allowing plastic flow in a zero (externally applied) stress state.

In view of the previous considerations, the total axial strain of a uni-axial element is written as the sum of a mechanical part (composed of an elastic and a plastic contribution), and of a hygro-expansive part:

$$\overline{\varepsilon}_i = \overline{\varepsilon}_i^e + \overline{\varepsilon}_i^p + \overline{\varepsilon}_i^h \tag{1}$$

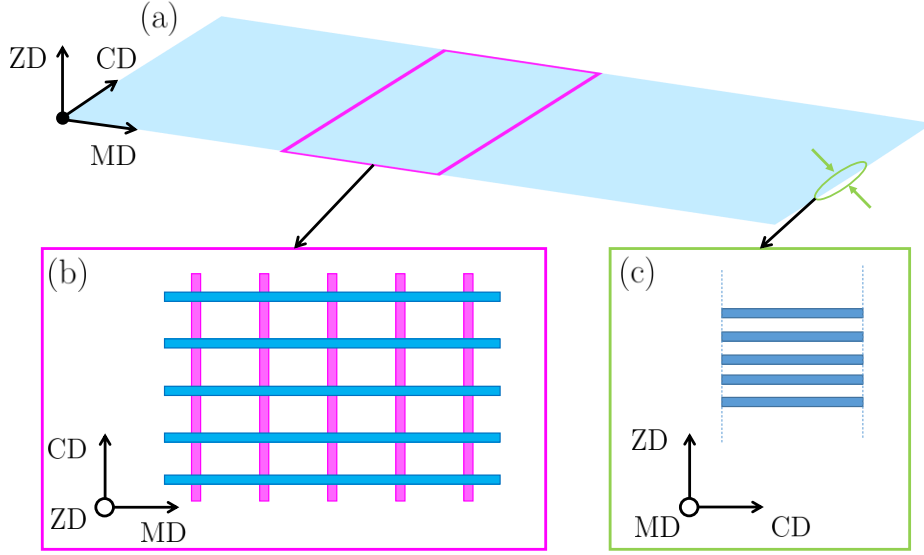

Figure 3: (a) Continuum paper sheet, and idealized discrete macro-scale geometry: (b) in-plane beam network and (c) extension in the out-of-plane direction.

where the hygro-expansive strain is given as

$$\overline{\varepsilon}_i^h = \overline{\beta}_i \chi \tag{2}$$

with $\chi$ the moisture content and $\overline{\beta}_i$ the macro-scale hygro-expansive coefficient. Here, and in all further relations, the subscript $i$ has to be read as $i = \mathrm{MD}, \mathrm{CD}$; a superimposed bar denotes macro-scale quantities.

The stress-strain relation is given by

$$\overline{\sigma}_i = \overline{E}_i(\overline{\varepsilon}_i - \overline{\varepsilon}_i^p - \overline{\varepsilon}_i^h) \tag{3}$$

where $\overline{E}_i(\chi)$ is the Young's modulus, which depends on the moisture content.

The yield function is given as

$$g(\overline{\sigma}_i, \overline{q}_i) = |\overline{\sigma}_i - \overline{q}_i| - \overline{\sigma}_i^y \leq 0 \tag{4}$$

where $\overline{q}_i$ and $\overline{\sigma}_i^y$ indicate the back stress and the yield stress, respectively. Note that $\overline{\sigma}_i^y$ is a function of $\chi$ only; $\overline{q}_i$ evolves as discussed below.

The associative flow rule is

$$\dot{\overline{\varepsilon}}_i^p = \dot{\gamma}\,\mathrm{sign}(\overline{\sigma}_i - \overline{q}_i) \tag{5}$$

with $\dot{\gamma} \geq 0$, such that the yield criterion is satisfied. The linear kinematic hardening law is written as

$$\dot{\overline{q}}_i = \overline{K}_i\,\dot{\overline{\varepsilon}}_i^p \tag{6}$$

with $\overline{K}_i(\chi)$ the moisture dependent hardening modulus. Note that, in the literature, more advanced hardening laws for paper have been presented (see e.g. [31]). For the sake of simplicity, it is assumed that linear kinematic hardening adequately captures the moisture effects on paper deformation.

A linear dependency of the yield stress on the moisture content is assumed as

$$\overline{\sigma}_i^y(\chi) = \overline{\sigma}_i^{y0}\left(1 - \frac{\chi}{\chi_0}\right) \tag{7}$$

where $\overline{\sigma}_i^{y0}$ is the yield stress at $\chi = 0$ and $\chi_0$ is the moisture content, equal in MD and CD, at which the yield stress vanishes. According to the experimental results reported in [20, 31], the Young's modulus and the hardening modulus are also linearly dependent on the moisture content as

$$\overline{E}_i(\chi) = \overline{E}_i^0 \left(1 - \frac{\chi}{\chi_E}\right) \tag{8}$$

$$\overline{K}_i(\chi) = \overline{K}_i^0 \left(1 - \frac{\chi}{\chi_K}\right) \tag{9}$$

Here, $\overline{E}_i^0$ and $\overline{K}_i^0$ are the Young's modulus and the hardening modulus at $\chi = 0$; $\chi_E$ and $\chi_K$ are the moisture contents, again equal in MD and CD, at which these moduli vanish.

## 3. Macro-scale numerical examples

*3.1. Parameters and considered geometries*

The material parameters have been taken from [20], where sheet level hygro-elasto-plastic properties have been calibrated to free expansion experiments. These data have been used to identify the material properties for the adopted discrete system, by enforcing equivalence between the response of a discrete elementary cell and of a corresponding continuum portion of the sheet. For details, the reader may refer to [20]. The properties of the discrete system are presented in Table 1.

| Parameter | Value | Unit | Parameter | Value | Unit | Parameter | Value | Unit |
|---|---|---|---|---|---|---|---|---|
| $\overline{\sigma}_{MD}^{y0}$ | $3.9 \cdot 10^1$ | MPa | $\overline{\sigma}_{CD}^{y0}$ | 9.4 | MPa | $\chi_0$ | $9.6 \cdot 10^{-2}$ | - |
| $\overline{E}_{MD}^0$ | $5.3 \cdot 10^3$ | MPa | $\overline{E}_{CD}^0$ | $1.6 \cdot 10^3$ | MPa | $\chi_E$ | $1.6 \cdot 10^{-1}$ | - |
| $\overline{K}_{MD}$ | $4.7 \cdot 10^3$ | MPa | $\overline{K}_{CD}$ | $4.8 \cdot 10^2$ | MPa | $\chi_K$ | $1.3 \cdot 10^{-1}$ | - |
| $\overline{\beta}_{MD}$ | $5.2 \cdot 10^{-2}$ | - | $\overline{\beta}_{CD}$ | $2.5 \cdot 10^{-1}$ | - | | | |

Table 1: Material parameters for the beams of the discrete system extracted from experimental data.

In all the considered numerical examples, the paper sheet is subjected to a moisture cycle in the range $\chi \in [\chi_D, \chi_W]$, with $\chi_D = 0.042$ the moisture content in the dry state and $\chi_W = 0.092$ that in the wet state (note that this value is well below the values of $\chi_0, \chi_E, \chi_D$, at which the material properties vanish). The total moisture variation considered is thus $\Delta\chi = \chi_W - \chi_D = 0.05$.

The decimal notation used here for the moisture content is totally equivalent to the conventionally employed percentage definition. For instance, $\chi_W = 0.092 = 9.2\%$.

The proposed phenomenological model can be used to capture the out-of-plane hygro-expansive response of paper induced by different moisture and constraining conditions. Different geometries and boundary conditions are therefore considered, which are illustrated in Figure 4. The problem in Figure 4(a) presents a paper strip whose length direction is parallel to the cross direction. The left edge of the sample is fully constrained, whereas the right edge is unconstrained. The strip is subjected to a wetting cycle, consisting of the initial dry state (1), the wet state (2) and the final dry state (3). Only the top surface is wetted; moisture penetrates for a thickness $h_w$. A piecewise constant moisture profile through the thickness is assumed; the moisture content does not vary in the plane. The dimensions are $a = 1$ mm, $h = 0.1$ mm. Figure 4(b) illustrates the same geometry and moisture profile of Figure 4(b). In this case, however, the sample is additionally placed on a solid surface which prevents bending during the moisture cycle. At the end of the drying stage, the paper strip is removed from the solid surface and it is free to deform out-of-plane. The problems illustrated in Figures 4(a) and 4(b) refer to the qualitative example discussed in Figure 1. The effect of an in-plane constraint is investigated through the example of Figure 4(c), in which during a uniform moisture cycle the paper sheet is mechanically constrained by two clamps, which are diagonally oriented with respect to MD and CD. Two cases are considered: (1) a narrow sample with $a = 5$mm, $b = 25$mm and (2) a wider sample with $a = 10$mm, $b = 20$mm.

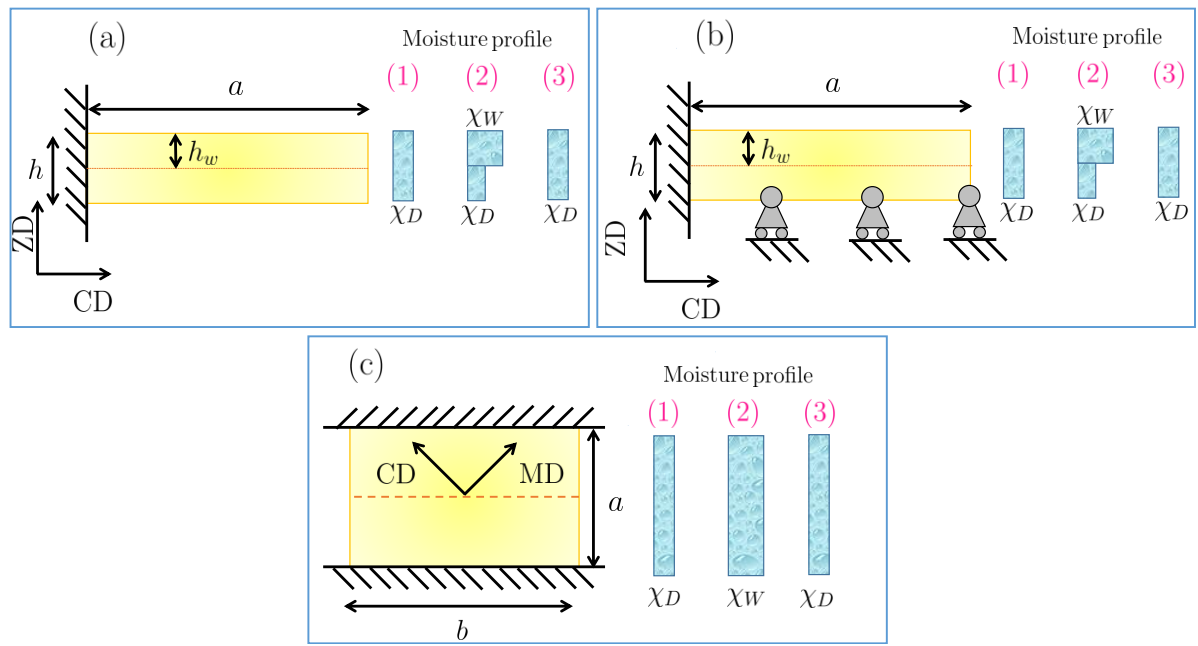

Figure 4: Geometries and boundary conditions considered in the macro-scale numerical examples: (a) unconstrained strip and (b) constrained strip, subjected to through the thickness moisture gradients; (c) paper sheet subjected to uniform moisture variation and in-plane constraint. The considered moisture cycles illustrate the initial dry state (1), the wet state (2) and the final dry state (3).

*3.2. Free bending due to moisture gradients through the thickness*

The example of Figure 4(a) serves to investigate the effects of moisture gradients through the thickness, in free expansion conditions. Moisture penetration in only half of the thickness of the paper strip is first considered. For the numerical solution of the problem, the thickness of the sheet is discretized into $n_l$ layers, with the condition that during deformation the right edge (section) remains straight. This condition, necessary to guarantee correspondence of the deformation of the discrete model with that of a continuous paper sheet, is enforced by tying constraints between the nodes on the right edge. For this particular boundary value problem, a further simplification can be made, by modelling the uni-axial layers through the thickness via truss elements, see Figure 3(c). The out-of-plane curvature is therefore calculated as a function of the relative displacement of the nodes on the right edge of the top and bottom trusses. Each truss is characterized by the CD constitutive behaviour described in Section 2.3.

Figure 5 shows the evolution during a moisture cycle of the normalized curvature (i.e. the curvature multiplied by the height $h$) and of the average along the thickness of the total strains of the layers. Note that the strains in the different layers vary linearly with their position along the thickness of the sheet; therefore, they can be reconstructed from the illustrated average value. The curves converge to each other for an increasing number of layers. In the wet state ($\Delta\chi = 0 \to 0.05$), the top region expands whereas the bottom region does not. The different deformations result in a downwards curvature of the strip, see Figure 5(a). In the drying stage ($\Delta\chi = 0.05 \to 0$), the hygro-expansive strain in the top region decreases, and likewise also the average strain and the curvature. For $n_l = 2$, the process is reversible due to the fact that the constraint on the right edge is not effective and, since in CD the material does not have an initial internal stress, as discussed in Section 2.1, all deformation can be accommodated by the curvature. The strip recovers its undeformed shape once the initial moisture content has been reached. For larger $n_l$, during the wetting stage, the piecewise constant moisture profile over the thickness combined with the fact that the right edge should remain straight results in stresses higher than the yield stress and therefore in a small amount of irreversible deformation. As a consequence, the maximum average total strain and curvature are between 10-35% less than for $n_l = 2$. The layers do not yield all at the same time; the strain profile shows therefore a slight curvature indicating this smooth transition. In the drying phase, there is a partial recovery of the plastic strain, resulting in a final negative average strain of approximately $\bar{\varepsilon}^{irr} = -1 \cdot 10^{-3}$. Note that the above results comply with the qualitative behaviour of the free strip shown in Figure 1.

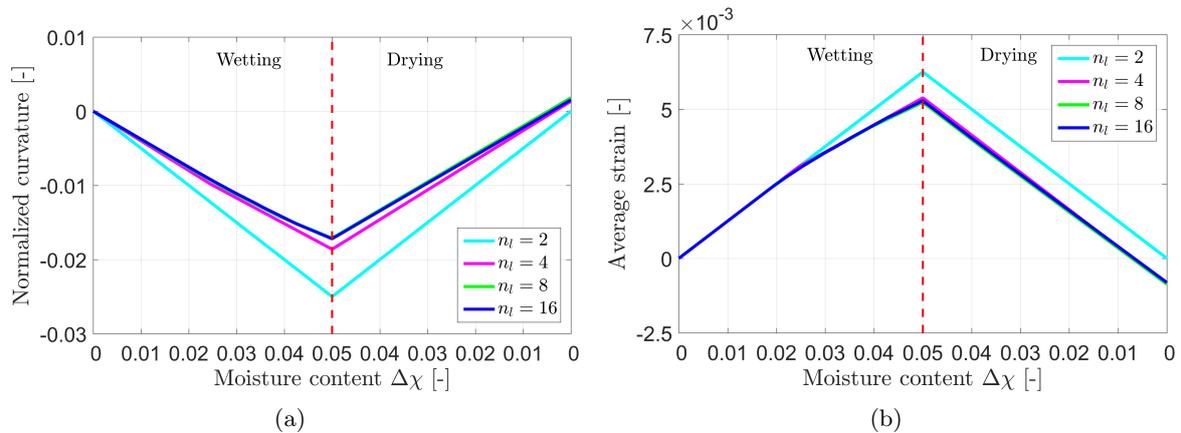

Figure 5: Normalized curvature in an unconstrained paper strip (a) and average total strain (b) as a function of the moisture content variation, for a moisture penetration of half of the thickness.

Figure 6(a) illustrates the internal stresses developed through the thickness of the paper strip in the wet state. These stresses have zero average, as the described process is overall stress free.

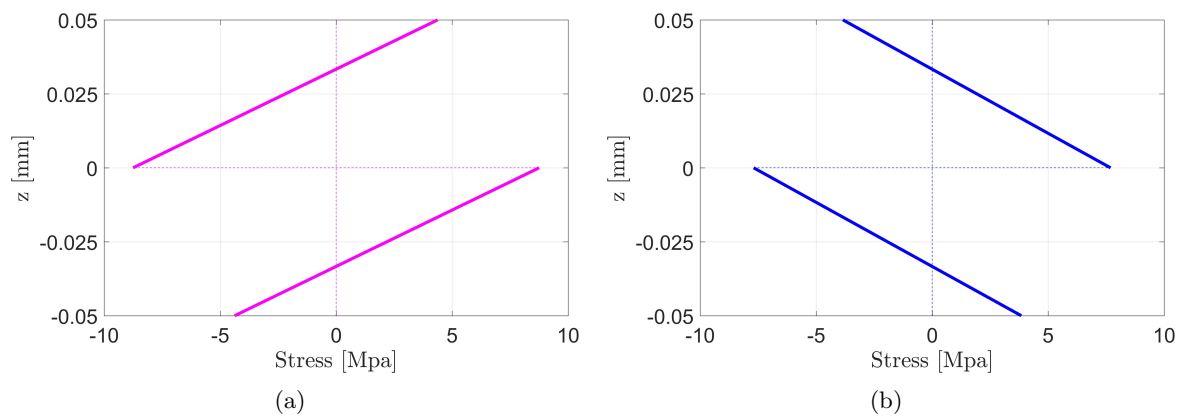

Figure 6: Internal stresses developed through the thickness of the sheet for a moisture penetration of half of the thickness: (a) in the wet state, for an unconstrained paper strip; (b) in the final dry state, for a paper strip that was constrained during the moisture cycle (see Section 3.3).

The influence of the normalized penetration depth, i.e. the ratio $h_w/h$ between the wet thickness and the total thickness, on the normalized curvature and on the average total strain is investigated in Figure 7. In this case, $n_l = 16$. Figure 7(a) shows the maximum total strain and curvature, i.e. in the wet state. The average total strain increases for an increasing penetration depth, because of the increasing contribution of the hygro-expansive strain in the wetted layers. The curvature initially increases for increasing penetration depths due to a higher driving force for bending. For larger penetrations, the curvature decreases as the contribution of the wet layers beneath the center line counters the effect of the layers above it. Figure 7(b) shows the final results after drying. The average total strain is always negative, and does not present a large variation as a function of the penetration depth. The normalized curvature shows a more complex profile. This depends on the condition that the section must remain straight upon wetting. Until about half of the wetting ratio, the curvature is positive due to the negative plastic strain in the top, wet layers. For higher penetration depths, positive plastic strain in some of the layers arises countering the effect of the other wet layers above the center line. The net normalized curvature becomes slightly negative. Finally, at high penetrations, approaching uniform wetting conditions, the normalized curvature and average strain decrease to zero again.

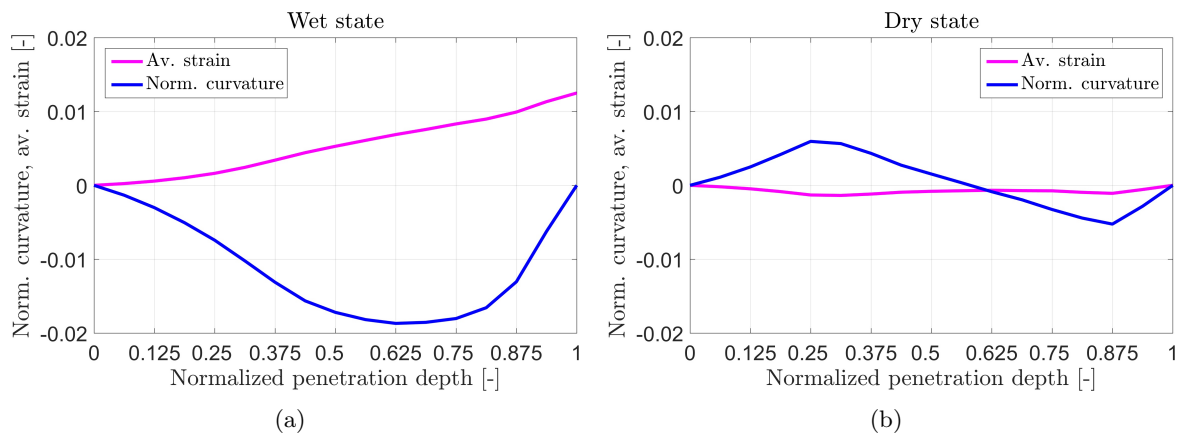

Figure 7: Normalized curvature and average total strain in an unconstrained paper strip as a function of the normalized penetration depth in the (a) wet and (b) dry state.

*3.3. Combined effect of moisture gradients through the thickness and applied mechanical constraints*

The combined effect of moisture variation through the thickness and constraining during the moisture cycle is studied by adopting the geometry and boundary conditions of Figure 4(b). A significantly different behaviour from the case illustrated in Section 3.2 is observed. Figure 8 shows (a) the normalized curvature and (b) the average total strain of the paper strip during the moisture cycle, in which the moisture penetrates in half of the thickness. The curvature stays zero until the release of the constraint. In the wetting stage ($\Delta\chi = 0 \rightarrow 0.05$), the differential expansion between the wet and the dry region cannot be accommodated by out-of-plane deformations, as the bending is restricted. This results in an internal compressive stress in the wet layers and a tensile stress in the dry layers. The yield stress decreases as moisture increases. The wet region therefore yields first and negative plastic strain occurs, which is observed in Figure 8(b) at $\Delta\chi = 0.018$ as a change of slope. The dry region remains elastic, as due to the lower moisture content the yield stress is higher.

In the drying stage ($\Delta\chi = 0.05 \rightarrow 0$), the hygro-expansive strain in the wet layer decreases. This results in elastic unloading and subsequent development of a tensile stress. The plastic strain starts to evolve again at $\Delta\chi = 0.042$ , leading to a slight kink in the curve of Figure 8(b). The negative plastic strain in the wet layers is partially released until the end of the moisture cycle. The kinematical constraint is removed at the end of the process, allowing the system to relax the internal stress which has developed during the cycle by bending upward, see Figure 8(a). For increasing $n_l$, while the average strain is constant, the results converge to the same final curvature. Note that the absolute value of the final curvature is much larger (about six times, for the converged result) compared to the case of unconstrained expansion, and it is about half of the value in the unconstrained wet state. Again, the results of this proposed simple model agree with the qualitative response of the constrained strip shown in Figure 1.

Note that a considerable amount of stress is generated (and preserved) in the layers during the constrained moisture cycle, as it can be seen in 6(b). This is captured by the plastic part of the model. This induced stress has a strong analogy with the initial dried-in strain developed in the production process, except for the fact that it is driven by moisture gradients through the thickness.

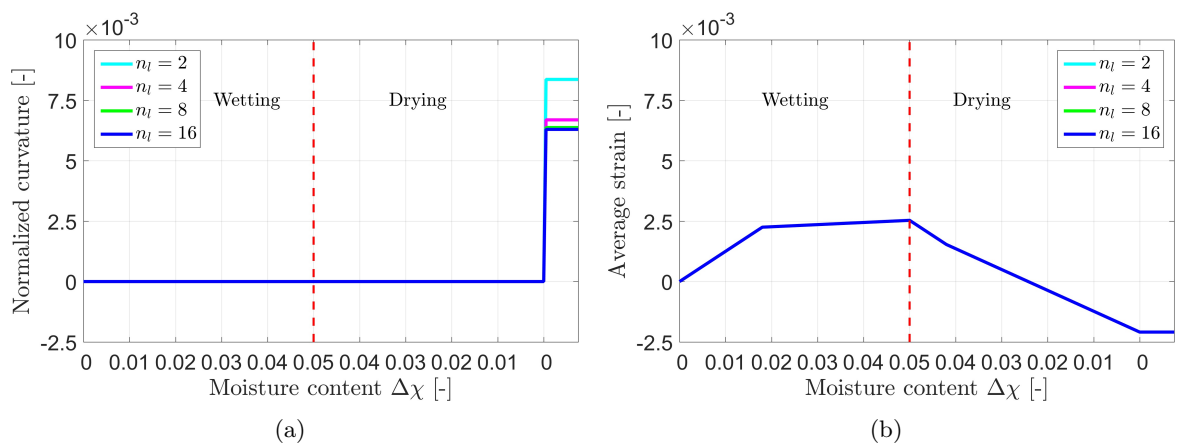

Figure 8: Normalized curvature (a) and average total strain (b) as a function of the moisture content variation, for a moisture penetration of half of the thickness in constrained expansion conditions.

The influence of the penetration depth on the out-of-plane deformation of the paper strip is finally investigated. Figure 9 shows the average total strain and normalized curvature as a function of the normalized penetration depth. In the wet state, shown in Figure 9(a), the average total strain obviously increases as a function of the penetration depth due to the larger hygro-expansive contribution, and the curvature is zero, as imposed by the constraint. In Figure 9(b), the average strain and normalized curvature are given in the dry state after releasing the constraint. The absolute values of both the average total strain and the normalized curvature initially increase for larger penetration depths. A larger number of wetted layers results in a larger final plastic strain. However, when the penetration depth increases further the stress decreases in the wet layers, triggering a decrease of the negative plastic strain and therefore of the absolute average strain and normalized curvature.

3.4. Buckling induced by in-plane kinematical constraint

The effect of in-plane mechanical constraints on the out-of-plane deformation of paper is discussed next, using the boundary conditions of Figure 4(c). Figure 10(top) presents the contour plots of the out-of-plane displacement in the fully wet state for the narrow sample (1) and the wide sample (2). Buckles arise along the diagonal direction between the clamps, where the diagonals with zero $z$-displacements between the buckles are parallel to MD and the "wave" of the buckles aligns in CD. This orientation is due to the fact that the hygro-expansion is significantly larger in CD than in MD. In Figure 10(bottom), the simulated results are compared with an experiment, in which the paper sample is positioned in a chamber with constant relative humidity of 97%. By removing the sample from the container it is exposed to ambient humidity, which is measured to vary between 54% and 56%. This has been chosen to represent the initial dry state.

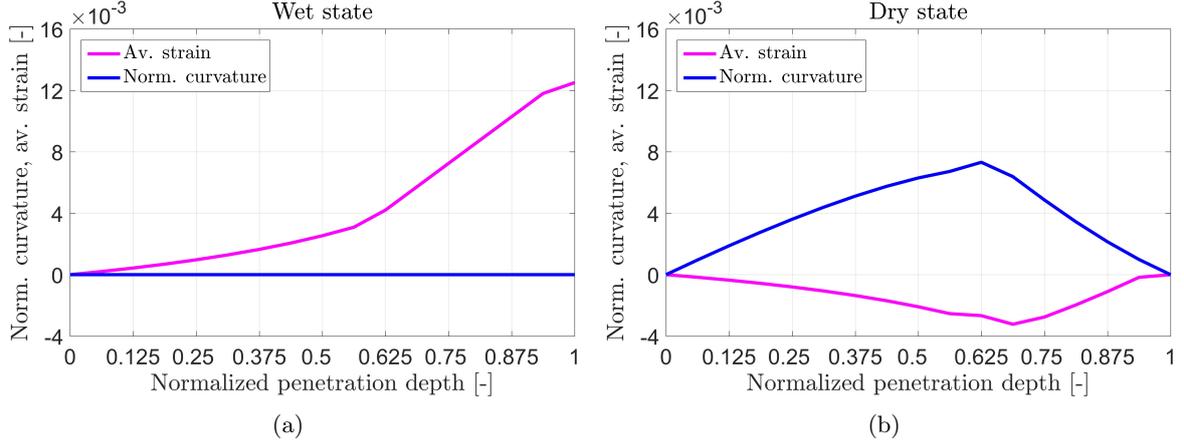

Figure 9: Normalized curvature and average total strain as a function of the normalized penetration depth (a) in the wet state in constrained expansion conditions and (b) in the dry state, after releasing the constraint.

The height profile is measured with an optical profilometer, both in the wet state and after drying. The measurements are performed three hours after changing the relative humidity. From the comparison between the experimental and numerical results, shown in Figure 10, an adequate qualitative agreement can be noticed.

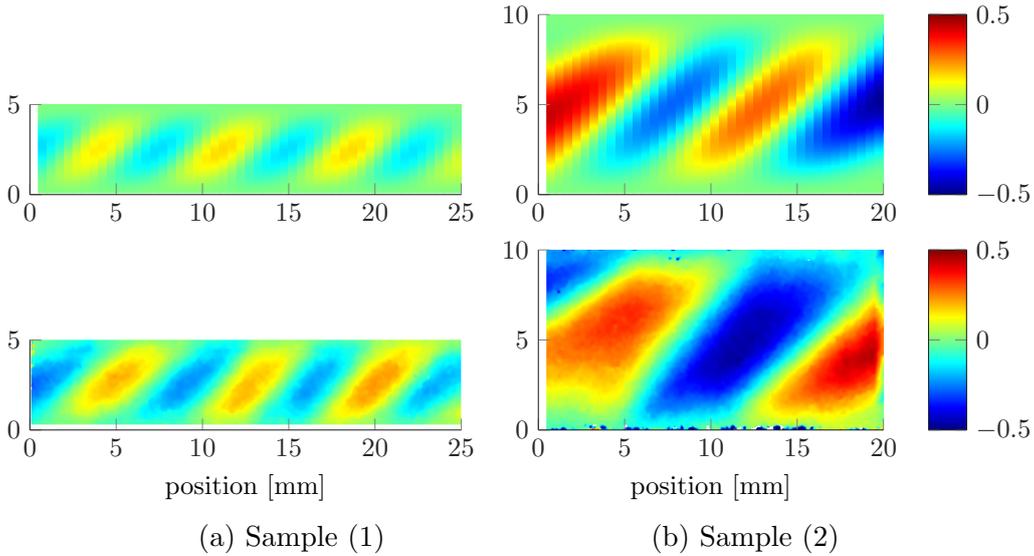

(a) Sample (1)  (b) Sample (2)

Figure 10: Contour plot of the out-of-plane displacement in the wet state: numerical (top) and experimental (bottom) height profiles in the wet state, for sample (1) and sample (2), in (a) and (b), respectively. The contour bar referring to the ZD displacement is expressed in mm.

Figure 11 quantitatively compares the height profiles along the center lines of the samples, i.e. the red dashed line in Figure 4(c). The black lines refer to the numerical results, while the coloured lines refer to the measurements. The simulations and the experiments reveal similar buckling wavelengths and amplitudes, particularly considering the scatter in the experimental data. In the narrow sample (1), the height profiles are characterized by a pattern containing multiple wavelengths, see Figure 11(a). In the wide sample (2), shown in Figure 11(b) only one wavelength is observed in the wet state. The wavelength is larger for the wide sample than for the narrow sample (approximately two times); the same trend is observed for the amplitude of the displacement. In fact, if the constraints span a larger distance, the center of the sample has more kinematical freedom and it assumes a larger wave shape to relax its stored energy.

In the narrow sample, the constraint enforces a smaller wavelength and amplitude. In the dry state, both samples show only a small out-of-plane deformation with wavelengths that are comparable to the wavelengths in the wet state, especially for sample (2). Quantitatively, the results of the simulations underestimate the wavelength by approximately 25% and the amplitude by 33%. Considering the simplicity of the model and the experimental uncertainty, this is a very satisfactory result.

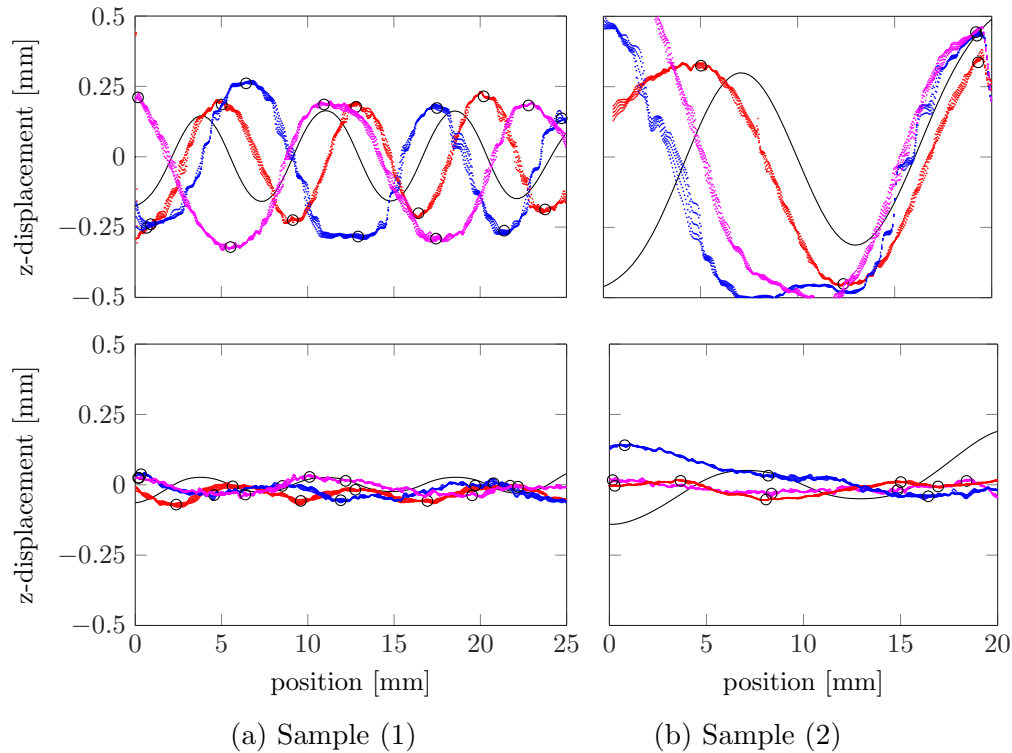

(a) Sample (1)  (b) Sample (2)

Figure 11: Height profiles along the center lines in the wet (top) and in the dry (bottom) state, for sample (1) and sample (2), in (a) and (b), respectively. The black curves refer to simulations, the coloured lines refer to three different experimental measurements.

## 4. Towards network level understanding of moisture induced deformations

### 4.1. Meso-scale model

The macro-scale phenomenological model proposed in Section 2 has been shown to be able to capture moisture induced out-of-plane deformations for different relevant examples. However, the fundamental physical mechanisms responsible for the sheet level behaviour of paper originate at the level of the underlying fibrous network. To understand the multi-scale nature of the hygro-mechanical response, a two dimensional periodic representative volume element of the fibrous network has been developed. The RVE is characterized by an edge length $L$ and area $Q = [0, L] \times [0, L]$. The fibres are assumed to be rectangular, with length $l$ and width $w$. In order to ensure that the RVE is representative of the effective material behaviour, the length of its edge is related to the fibre length via $L = 2l$ [27, 32]. Defining the mean areal coverage $\bar{c}$ as the ratio between the total area occupied by the fibres and the area of the RVE ($\bar{c} = A_f/Q$), the number of fibres $n_f$ within a network of given coverage is $n_f = (\bar{c}\, Q)/(lw)$.

The coverage can be interpreted as a measure of the effective number of fibre layers within the network. In each point in the plane, the stiffnesses of the individual layers through the thickness are added and the result is divided by the nominal sheet thickness $t\bar{c}$, with $t$ the thickness of a single fibre. This implies that all fibres that overlap in a certain point of the domain are considered to be perfectly bonded to each other. Essentially, the thickness direction is collapsed into a plane.

A uniform random point field defines the positions of the centre of each fibre within the domain $Q$. A wrapped Cauchy orientation distribution probability density function is used to represent the fibres' orientation [33]:

$$f(\theta) = \frac{1}{\pi} \frac{1 - q^2}{1 + q^2 - 2q\cos(2\theta)} \qquad (10)$$

where $-\pi/2 < \theta \leq \pi/2$ is the angle between the fibre axis and the machine direction and $0 \leq q < 1$ is a measure of the anisotropy of the network orientation.

Periodicity of the RVE is obtained by cutting fibre segments that exit the RVE boundaries and periodically copying them to the opposite edges.

Contrary to the macroscopic model adopted in the previous sections, the present network analysis is restricted to hygro-elastic behaviour. This is sufficient to identify the key mechanisms at the meso-structural level that influence macroscopic moisture induced deformation. Moreover, the influence of the moisture content on the elastic properties of the fibres is neglected. The dried-in stress and strain can be incorporated along the lines suggested in Reference [26] for a geometrically simpler model.

In view of the above considerations, a transversely isotropic hygro-elastic constitutive law is used to model the response of an individual fibre [34, 35]. Contrary to many works in the literature, where beam or truss approximations are used, the fibres are here considered as fully two dimensional solids. The effect of moisture on the mechanics of the network can be captured only by including both the longitudinal and the transverse response of a single fibre. This enables to analyse the interplay in the bonding regions between hygro-expansion and mechanics due to the different longitudinal and transverse properties of fibres. Full strain compatibility (i.e. perfect bonding) between the fibre layers that compose the bonds is assumed (for details, see [27]).

The network geometry is discretized and solved numerically as follows. A regular grid of $n_e$ square (finite) elements is constructed on the RVE. The element edge is $l_e = w/\xi$, where $\xi \geq 1$ is an integer. An element $e$ is considered part of a given fibre if its geometrical center lies inside of the fibre area, $l \times w$. The fibre boundaries as a result are approximated by zigzag edges. The proposed approximation may lead to local stress and strain concentrations at the fibres boundaries. However, for sufficiently large values of $\xi$ and for the hygro-elastic case considered here, this influences only minimally the deformation, which is relevant to understand how meso-structural mechanisms govern the overall response. In this case, $\xi$ has been set to five, i.e. on average five elements across the width of the fibre have been used. This approximation is sufficiently accurate: the error in the estimate of the overall hygro-expansive properties compared to a much larger value of $\xi$ ($\xi = 20$) is lower than 2%. Bilinear quadrilateral finite elements have been used. Periodic boundary conditions have been applied. The network generation and discretization and the finite element solution of the hygro-elastic problem have been implemented in an lab-developed MATLAB code.

### 4.2. Meso-scale response
#### 4.2.1. Input parameters and considered geometries

The material properties for the meso-structural level are taken from the literature [5, 34, 36]; their values are collected in Table 2.

For the network simulations, two network coverages have been considered: $\bar{c} = 1$ and $\bar{c} = 5$, corresponding to a sparse and a dense network, respectively. Additionally, for each coverage, two different orientation distributions have been used: $q = 0$ (uniformly distributed network) and $q = 0.5$ (highly oriented network). All the considered fibrous networks are subjected to a uniform moisture content variation $\chi$ in free expansion conditions. Note that since the problem is linear, all subsequent results will be normalized by $\chi$.

| Parameter | Value | Unit |
|---|---|---|
| $E_\ell$ | 35 | GPa |
| $E_t$ | 5.83 | GPa |
| $G_{\ell t}$ | 3.5 | GPa |
| $\nu_{\ell t}$ | 0.3 | - |
| $\nu_{t\ell}$ | 0.05 | - |
| $\beta_\ell$ | 0.03 | - |
| $\beta_t$ | 0.6 | - |

Table 2: Material parameters used in the fibre constitutive model, specified with respect to the fibre local coordinate system $(\ell, t)$, where the $\ell$-axis is coincident with the fibre axis.

4.2.2. Local network fields

The hygro-expansive response of the fibrous networks considered in the analysis is investigated in Figure 12. Figure 12(a) and 12(c) refer to coverage $\bar{c} = 1$ and $\bar{c} = 5$, respectively, with a uniform fibre orientation ($q = 0$). The deformation of the two networks with the same coverages but with an anisotropic fibre orientation ($q = 0.5$) is illustrated in Figure 12(b) and 12(d). Black lines show the fibres in the undeformed configuration, magenta lines refer to the deformed configuration. In the Figures, the $x-$ and $y-$ axes correspond to MD and CD, respectively. Note that for $\bar{c} = 5$, the network is extremely dense and essentially behaves as a heterogeneous continuum, as it can be also seen in Figure 13 .

The overall expansion is essentially governed by the hygro-expansion of the bonding regions. This due to the fact that the free-standing parts of the fibres i) swell only little in the longitudinal direction and ii) their lateral swelling is not transferred to the network and thus does not contribute. This explains why the high coverage networks with a larger number of bonds have a higher resulting expansion than the sparse networks. Moreover, a strong influence of the fibre orientation appears in the network expansion. As the anisotropy of the network increases, see Figure12(b) and 12(d), a larger expansion in CD occurs. This is due to the fact that, for an anisotropic orientation distribution, the bonds are composed on average by more fibres oriented along MD. The (larger) transverse expansion of the individual fibre contributes thus more to the resulting bond expansion in CD, increasing the overall CD deformation. This effect can be quantified by computing the average strain of all the bonds in the network and comparing the MD and CD components. The results are collected in Table 3(top). Whereas for the isotropic networks the average bond strain components are comparable, there is a strong predominance of the bond strain in CD in the case of anisotropic orientation.

The above observations are confirmed by the resulting overall hygro-expansive coefficients of the networks, shown in Table 3(bottom). The obtained values are close to the bond average strains - see Table 3(top)- supporting the relevance of the bond response to the overall hygro-expansion. Moreover, the computed hygro-expansive coefficients are in the same order of magnitude of the parameters of the macro-scale examples, particularly in the case of the network with coverage $\bar{c} = 5$ and $q = 0.5$ that are realistic values for actual printing paper, despite the strong assumption made on the full kinematical compatibility in the bonding regions. These findings are therefore relevant to understand the origin of the macroscopic anisotropic hygro-expansive coefficients, which influence the out-of-plane sheet-level response: maximum curvature and waves along the cross direction, shown in Sections 3.2-3.3 and in

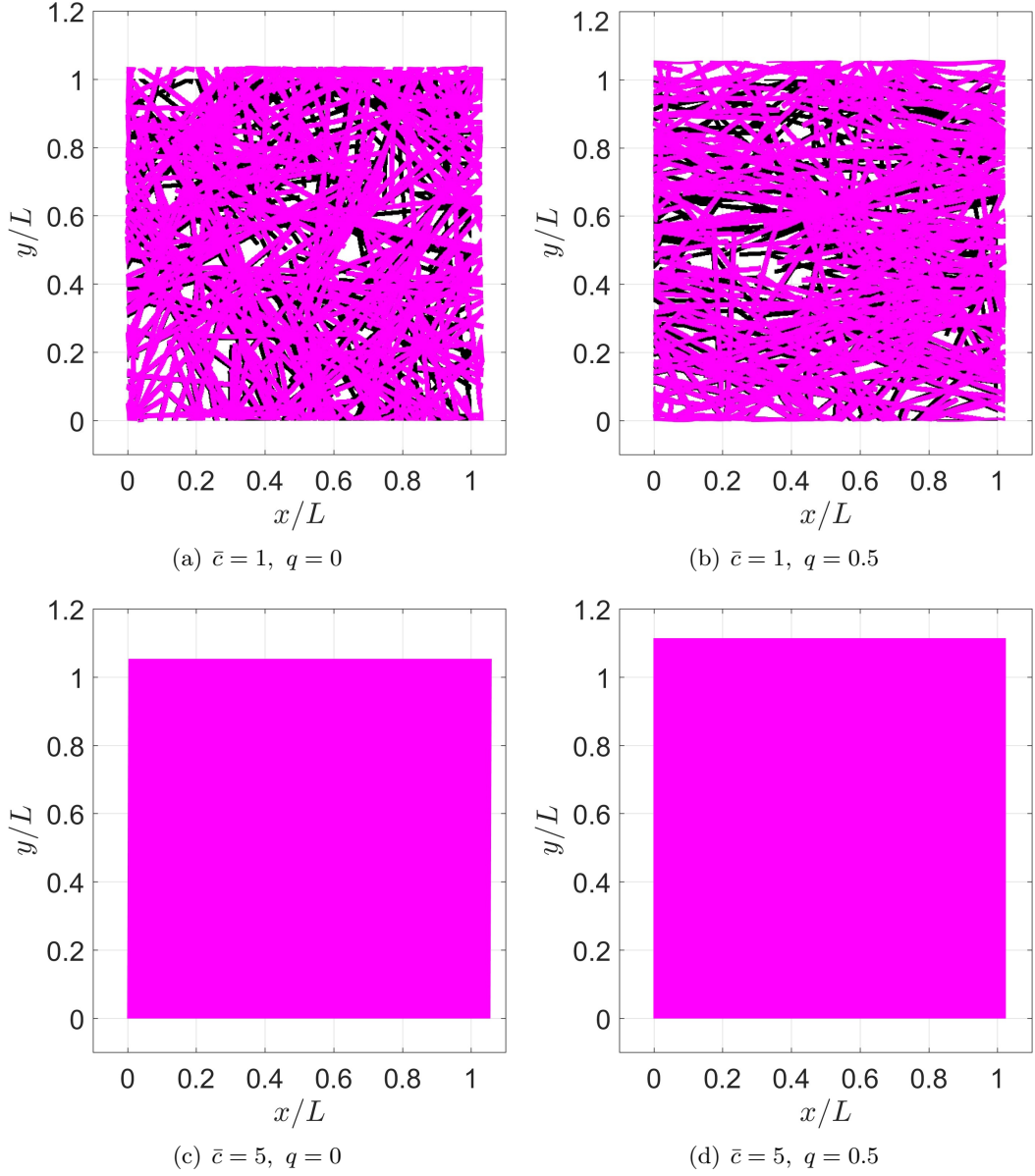

Figure 12: Deformed configuration (magenta lines) $\mathbf{u}/\chi$ due to free expansion for networks of coverages $\bar{c} = 1$ (a)-(b) and $\bar{c} = 5$ (c)-(d), for isotropic (a)-(c) and anistropic (b)-(d) fibre orientation.

Section 3.4, respectively.

| Average bond strain | $\bar{c} = 1$ | | $\bar{c} = 5$ | |
|---|---|---|---|---|
| | $q = 0$ | $q = 0.5$ | $q = 0$ | $q = 0.5$ |
| $\bar{\varepsilon}^b_{\mathrm{MD}}$ | 0.0801 | 0.0423 | 0.1102 | 0.0569 |
| $\bar{\varepsilon}^b_{\mathrm{CD}}$ | 0.0830 | 0.1203 | 0.1207 | 0.2603 |

| Hygro-expansion coefficients | $\bar{c} = 1$ | | $\bar{c} = 5$ | |
|---|---|---|---|---|
| | $q = 0$ | $q = 0.5$ | $q = 0$ | $q = 0.5$ |
| $\bar{\beta}_{MD}$ | 0.0808 | 0.0491 | 0.1205 | 0.0536 |
| $\bar{\beta}_{CD}$ | 0.0814 | 0.1281 | 0.1265 | 0.2603 |

Table 3: Overall hygro-expansive coefficients computed from the fibrous paper networks.

The stress distribution is shown for the two networks with anisotropic orientation, $\bar{c} = 1$ in Figure 13(a)-(b) and $\bar{c} = 5$ in Figure 13(c)-(d). The stresses in the bonds have to be interpreted as the average through the thickness of the stresses within the single fibre layers. Note first that the peak values of the stress are larger in the sparse network ($\bar{c} = 1$) than in the dense one ($\bar{c} = 5$). This is due to the fact that a denser network is more homogeneous and the stresses are therefore more uniformly distributed. Moreover, the stresses in the free-standing fibre segments are very low, as here only the expansive contribution to deformation plays a role. As discussed earlier, the misorientation of the different fibres and the consequent interplay between longitudinal and transverse properties within the bonds creates higher fluctuations in the strain distribution. Consequently, although the overall stress of the network is zero, internal stresses arise in the bonds. This can trigger local instabilities, for instance microbuckling of the compressed fibres within the bonds [37]. Note finally that the stresses are more localized along the cross direction than along the machine direction, triggered by the stronger internal constraint in CD with respect to MD.

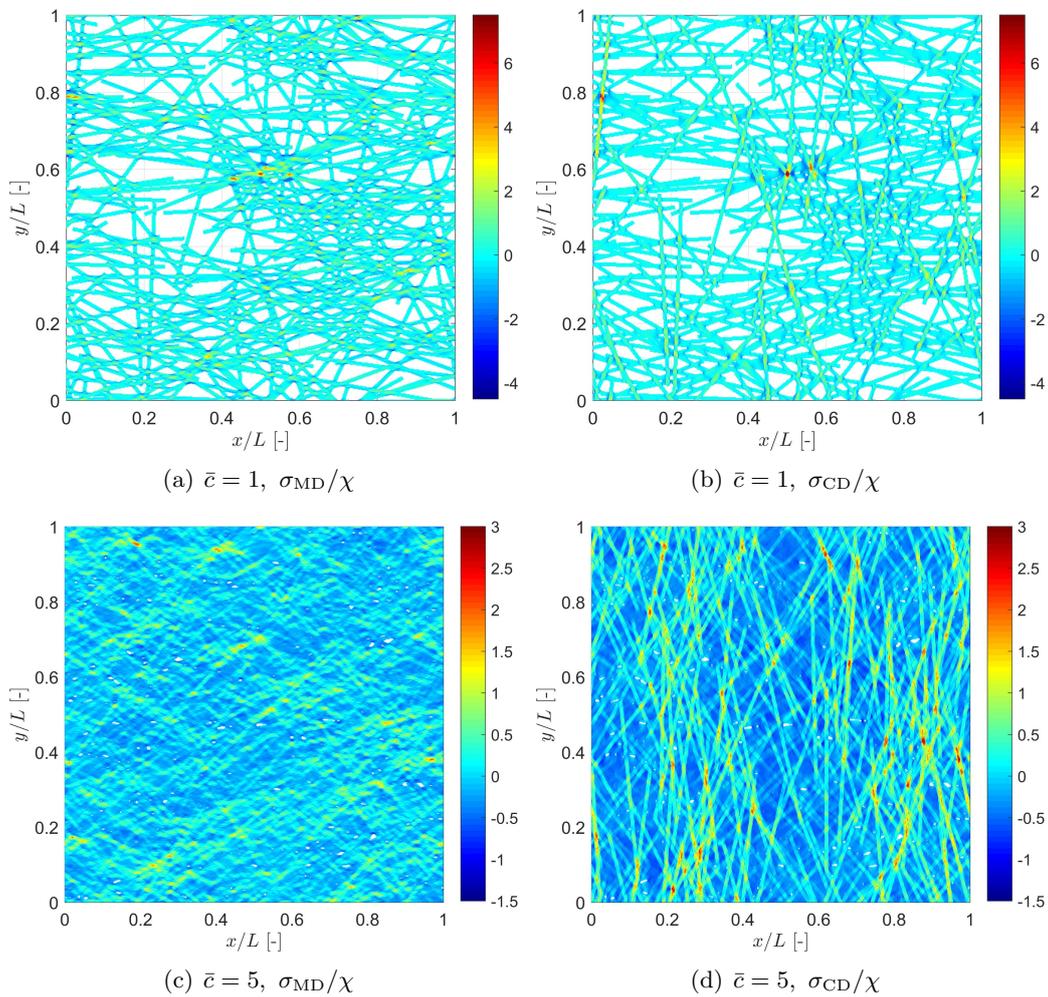

(a) $\bar{c} = 1$, $\sigma_{\text{MD}}/\chi$     (b) $\bar{c} = 1$, $\sigma_{\text{CD}}/\chi$

(c) $\bar{c} = 5$, $\sigma_{\text{MD}}/\chi$     (d) $\bar{c} = 5$, $\sigma_{\text{CD}}/\chi$

Figure 13: Stress distribution per unit moisture content $\boldsymbol{\sigma}/\chi$ (in GPa) for the networks of coverage $\bar{c} = 1$(a)-(b) and $\bar{c} = 5$(c)-(d), with anisotropic fibre distribution ($q = 0.5$).

## 5. Conclusions

The main focus of this paper was to explore the relation between moisture content variations and out-of-plane deformations in paper sheets.

A macroscopic phenomenological model for the hygro-mechanical response of paper has been proposed for this purpose. This model allows to describe moisture induced dimensional instability, incorporating the relevant effect of the release of dried-in strains induced during the production process. Despite its relative simplicity, the model has nevertheless a good predictive capability, demonstrated by the computed out-of-plane deformations resulting from different combinations of moisture distributions and mechanical constraints. In all the considered cases, the anisotropic sheet-level hygro-mechanical properties are a key factor governing potential instability phenomena. The effects of moisture gradients through the thickness have been investigated on a paper strip taken along the cross direction, in which the hygro-expansion is largest. Out-of-plane bending of the sheet occurs, with curvature magnitude and direction that are strongly influenced by possible constraining conditions during the moisture cycle. Small residual downwards curvature emerges in free expansion conditions, while a considerably higher (up to a factor of six) upwards curvature is revealed if the moisture cycle is combined to mechanical constraints. This latter effect depends on the stresses that are dried-in during the process. The obtained results comply with those of a qualitative experiment. In the case of uniform in-plane moisture variations in presence of mechanical constraints, the paper sheet buckles. The comparison between the numerical results and a consistent experimental test shows that the model captures the major aspects of the deformation. The experimental and simulated buckling patterns, whose direction depends on the anisotropic hygro-expansion, are in qualitative agreement. A semi-quantitative agreement has been found in the values of wavelengths and amplitudes. The obtained responses are thus adequate; however, additional experiments are needed for a better quantitative validation the proposed model. This will be the subject of future work.

The macroscopic response is ultimately controlled by a cascade of events occurring at the level of the underlying fibrous network. The origin of the paper macroscopic response has been investigated by developing a two dimensional representative volume element of the fibrous network, which accounts for all the relevant meso-structural features (single fibres and bond hygro-mechanical properties, fibres geometry and orientation...). Local deformations and stresses have been analysed by subjecting several fibrous networks to a uniform moisture content variation. The interplay between the expansion of the inter-fibre bonding regions and the fibre orientation distribution essentially determines the sheet-level anisotropic hygro-expansion, which has been already shown to critically influence macroscopic instabilities.

This work can be extended to a meso-scale three dimensional description, which would allow to represent the geometry more accurately and to investigate additional mechanisms at the network level (e.g. micro-buckling or fibre twisting), which further influence the sheet-scale hygro-expansive behaviour.

## References

[1] S. Douezan, M. Wyart, F. Brochard-Wyart, D. Cuvelier, Curling instability induced by swelling, Soft Matter 7 (2011) 1506–1511. doi:10.1039/C0SM00189A.

[2] A. Kulachenko, P. Gradin, T. Uesaka, Basic mechanisms of fluting formation and retention in paper, Mechanics of Materials 39 (7) (2007) 643 – 663. doi:http://dx.doi.org/10.1016/j.mechmat.2006.10.002.
URL http://www.sciencedirect.com/science/article/pii/S0167663606001293

[3] S. J. Hashemi, W. J. M. Douglas, Moisture nonuniformity in drying paper: Measurement and relation to process parameters, Drying Technology 21 (2) (2003) 329–347. arXiv:http://dx.doi.org/10.1081/DRT-120017754, doi:10.1081/DRT-120017754.
URL http://dx.doi.org/10.1081/DRT-120017754

[4] S. Gepp, J. Örtegren, J.-E. Hägglund, E. Alfthan, Measuring cockling on-line in high speed inkjet printing, International Conference on Digital Printing Technologies (2009) 521–523.

[5] K. Niskanen, Paper physics, Fapet Oy Helsinki, Finland, 1998.

[6] P. A. Larsson, L. Wagberg, Influence of fibre-fibre joint properties on the dimensional stability of paper, Cellulose 15 (4) (2008) 515–525. `doi:10.1007/s10570-008-9203-y`.

[7] C. Marulier, P. J. J. Dumont, L. Orgéas, S. Rolland du Roscoat, D. Caillerie, 3D analysis of paper microstructures at the scale of fibres and bonds, Cellulose 22 (3) (2015) 1517–1539. `doi:10.1007/s10570-015-0610-6`.
URL `http://dx.doi.org/10.1007/s10570-015-0610-6`

[8] P. Mäkelä, S. Östlund, Orthotropic elastic-plastic material model for paper materials, International Journal of Solids and Structures 40 (21) (2003) 5599 – 5620. `doi:http://dx.doi.org/10.1016/S0020-7683(03)00318-4`.
URL `http://www.sciencedirect.com/science/article/pii/S0020768303003184`

[9] Q. S. Xia, M. C. Boyce, D. M. Parks, A constitutive model for the anisotropic elastic-plastic deformation of paper and paperboard, International Journal of Solids and Structures 39 (15) (2002) 4053 – 4071. `doi:http://dx.doi.org/10.1016/S0020-7683(02)00238-X`.
URL `http://www.sciencedirect.com/science/article/pii/S002076830200238X`

[10] M. Wallmeier, E. Linvill, M. Hauptmann, J.-P. Majschak, S. Östlund, Explicit FEM analysis of the deep drawing of paperboard, Mechanics of Materials 89 (2015) 202 – 215. `doi:http://dx.doi.org/10.1016/j.mechmat.2015.06.014`.
URL `http://www.sciencedirect.com/science/article/pii/S0167663615001441`

[11] E. Borgqvist, M. Wallin, M. Ristinmaa, J. Tryding, An anisotropic in-plane and out-of-plane elasto-plastic continuum model for paperboard, Composite Structures 126 (2015) 184 – 195. `doi:http://dx.doi.org/10.1016/j.compstruct.2015.02.067`.
URL `http://www.sciencedirect.com/science/article/pii/S0263822315001555`

[12] Y. Li, S. E. Stapleton, S. Reese, J.-W. Simon, Anisotropic elastic-plastic deformation of paper: In-plane model, International Journal of Solids and Structures (2016) –`doi:http://dx.doi.org/10.1016/j.ijsolstr.2016.08.024`.
URL `http://www.sciencedirect.com/science/article/pii/S0020768316302372`

[13] T. Uesaka, K. Murakami, R. Imamura, Two-dimensional linear viscoelasticity of paper, Wood Science and Technology 14 (2) (1980) 131–141. `doi:10.1007/BF00584042`.
URL `http://dx.doi.org/10.1007/BF00584042`

[14] W. Lu, L. A. Carlsson, Influence of viscoelastic behavior on curl of paper, Mechanics of Time-Dependent Materials 5 (1) (2001) 79–100. `doi:10.1023/A:1009895419026`.
URL `http://dx.doi.org/10.1023/A:1009895419026`

[15] O. Girlanda, D. D. Tjahjanto, S. Östlund, L. E. Schmidt, On the transient out-of-plane behaviour of high-density cellulose-based fibre mats, Journal of Materials Science 51 (17) (2016) 8131–8138. `doi:10.1007/s10853-016-0083-5`.
URL `http://dx.doi.org/10.1007/s10853-016-0083-5`

[16] D. Tjahjanto, O. Girlanda, S. Östlund, Anisotropic viscoelastic-viscoplastic continuum model for high-density cellulose-based materials, Journal of the Mechanics and Physics of

Solids 84 (2015) 1 – 20. `doi:http://dx.doi.org/10.1016/j.jmps.2015.07.002`.
URL `http://www.sciencedirect.com/science/article/pii/S0022509615300077`

[17] D. Roylance, Viscoelastic properties of paper, Fiber Science and Technology 13 (1980) 411–421.

[18] J. Lif, S. Östlund, C. Fellers, Applicability of anisotropic viscoelasticity of paper at small deformations, Mechanics of Time-Dependent Materials 2 (3) (1998) 245–267. `doi:10.1023/A:1009818022865`.
URL `http://dx.doi.org/10.1023/A:1009818022865`

[19] A.-L. Erkkilä, T. Leppänen, J. Hämäläinen, T. Tuovinen, Hygro-elasto-plastic model for planar orthotropic material, International Journal of Solids and Structures 62 (2015) 66 – 80. `doi:10.1016/j.ijsolstr.2015.02.001`.

[20] C. G. V. der Sman, E. Bosco, R. H. J. Peerlings, A model for moisture-induced dimensional instability in printing paper, Accepted for publication in Nordic Pulp and Paper Research Journal.

[21] M. Ostoja-Starzewski, Random field models of heterogeneous materials, International Journal of Solids and Structures 35 (19) (1998) 2429 – 2455. `doi:http://dx.doi.org/10.1016/S0020-7683(97)00144-3`.
URL `http://www.sciencedirect.com/science/article/pii/S0020768397001443`

[22] C. A. Bronkhorst, Modelling paper as a two-dimensional elastic-plastic stochastic network, International Journal of Solids and Structures 40 (20) (2003) 5441 – 5454. `doi:10.1016/S0020-7683(03)00281-6`.

[23] M. K. Ramasubramanian, Y. Wang, A computational micromechanics constitutive model for the unloading behavior of paper, International Journal of Solids and Structures 44 (2223) (2007) 7615 – 7632.

[24] J. Strömbro, P. Gudmundson, Mechano-sorptive creep under compressive loading a micromechanical model, International Journal of Solids and Structures 45 (9) (2008) 2420 – 2450. `doi:10.1016/j.ijsolstr.2007.12.002`.

[25] A. Kulachenko, T. Uesaka, Direct simulations of fiber network deformation and failure, Mechanics of Materials 51 (0) (2012) 1 – 14.

[26] E. Bosco, R. H. J. Peerlings, M. G. D. Geers, Explaining irreversible hygroscopic strains in paper: a multi-scale modelling study on the role of fibre activation and micro-compressions, Mechanics of Materials 91, Part 1 (2015) 76 – 94. `doi:10.1016/j.mechmat.2015.07.009`.

[27] E. Bosco, R. H. J. Peerlings, M. G. D. Geers, Asymptotic homogenization of the hygro-thermo-mechanical properties of fibrous networks., arXiv preprint arXiv:1609.07622.

[28] E. Bosco, R. H. J. Peerlings, M. G. D. Geers, Hygro-mechanical properties of paper fibrous networks through asymptotic homogenization and comparison with idealized models., arXiv preprint arXiv:1609.07623.

[29] E. Bosco, M. V. Bastawrous, R. H. J. Peerlings, J. P. H. Hoefnagels, M. G. D. Geers, Bridging network properties to the effective hygro-expansive behaviour of paper: experiments and modelling, Philosophical Magazine 95 (28-30) (2015) 3385–3401. `doi:10.1080/14786435.2015.1033487`.

[30] T. Uesaka, C. Moss, Y. Nanri, The characterization of hygroexpansivity of paper, Journal of Pulp and Paper Science 18 (1) (1992) J11 – J16.

[31] A.-L. Erkkilä, T. Leppänen, J. Hämäläinen, Empirical plasticity models applied for paper sheets having different anisotropy and dry solids content levels, International Journal of Solids and Structures 50 (1415) (2013) 2151 – 2179. `doi:http://dx.doi.org/10.1016/j.ijsolstr.2013.03.004`.
URL `http://www.sciencedirect.com/science/article/pii/S0020768313000954`

[32] H. Hatami-Marbini, R. C. Picu, Heterogeneous long-range correlated deformation of semiflexible random fiber networks, Phys. Rev. E 80 (2009) 046703. `doi:10.1103/PhysRevE.80.046703`.

[33] H. Cox, The elasticity and strength of paper and other fibrous materials, British Journal of Applied Physics 3 (3) (1952) 72.

[34] A. Bergander, L. Salmén, Cell wall properties and their effects on the mechanical properties of fibers, Journal of Materials Science 37 (1) (2002) 151–156. `doi:10.1023/A:1013115925679`.

[35] S. Borodulina, A. Kulachenko, D. Tjahjanto, Constitutive modeling of a paper fiber in cyclic loading applications, Computational Materials Science 110 (2015) 227–240. `doi:10.1016/j.commatsci.2015.08.039`.

[36] C. Sellén, P. Isaksson, A mechanical model for dimensional instability in moisture-sensitive fiber networks, Journal of Composite Materials 48 (3) (2014) 277–289. `doi:10.1177/0021998312470576`.

[37] H. W. Haslach, A model for drying-induced microcompressions in paper: Buckling in the interfiber bonds, Composites Part B: Engineering 27 (1) (1996) 25 – 33. `doi:http://dx.doi.org/10.1016/1359-8368(95)00003-8`.
URL `http://www.sciencedirect.com/science/article/pii/1359836895000038`